# Spatial Measures of Urban Systems: from Entropy to Fractal Dimension

Yanguang Chen; Linshan Huang

(Department of Geography, College of Urban and Environmental Sciences, Peking University, Beijing 100871, P.R. China. E-mail: chenyg@pku.edu.cn)

**Abstract**: A type of fractal dimension definition is based on the generalized entropy function. Both entropy and fractal dimension can be employed to characterize complex spatial systems such as cities and regions. Despite the inherent connect between entropy and fractal dimension, they have different application scopes and directions in urban studies. This paper focuses on exploring how to convert entropy measurement into fractal dimension for the spatial analysis of scale-free urban phenomena using ideas from scaling. Urban systems proved to be random prefractal and multifractals systems. The entropy of fractal cities bears two typical properties. One is the scale dependence. Entropy values of urban systems always depend on the scales of spatial measurement. The other is entropy conservation. Different fractal parts bear the same entropy value. Thus entropy cannot reflect the spatial heterogeneity of fractal cities in theory. If we convert the generalized entropy into multifractal spectrums, the problems of scale dependence and entropy homogeneity can be solved to a degree for urban spatial analysis. The essence of scale dependence is the scaling in cities, and the spatial heterogeneity of cities can be characterized by multifractal scaling. This study may be helpful for the students to describe and understand spatial complexity of cities.

# 1. Introduction

Urban systems indicate both cities as systems and the systems of cities. A city as a system is concept of individual city and belongs to intraurban geography, and a system of cities is a concept of urban network and belongs to interurban geography (De Blij and Muller, 1997). Both cities and systems of cities proved to be self-organizing complex spatial systems (Allen, 1997; Portugali, 2011; Wilson, 2000). Complex systems can be described with entropy (Bar-Yam, 2004; Batty *et al*, 2014; Cramer, 1993), including Boltzmann's macro and micro state entropy, Shannon's information, and Renyi's generalized entropy. Unfortunately, in many cases, entropy values depend on the scale of measurement (Chen *et al*, 2017). If we study a city as a system, the entropy values in different years may be incomparable; if we research a system of cities, the entropy values of different cities may be incomparable due to the difference of resolution ratios of remote sensing images. Scale dependence of entropy influences the effect of spatial analysis for urban systems. One of method to solve this problem is to replace entropy with fractal dimension in light of the inherent relationship between entropy and fractal dimension (Chen *et al*, 2017; Ryabko, 1986; Stanley and Meakin, 1988).

Fractal dimension is the basic parameter for describing self-similar patterns and processes. A fractal has three typical properties: scaling law, fractional dimension, and entropy conversation law (Chen, 2016). Scaling law implies the scale dependence of spatial measurement of fractal systems, and entropy conversation suggests that the spatial heterogeneity cannot be effectively reflected by entropy values. On the other hand, there are two ways to define fractal dimension. One is based on entropy function, and the other is based on correlation function. The two ways are equivalent to one another, but the angles of view are different. Based on entropy functions, the models are expressed as logarithmic functions or exponential functions, while based on correlation function, the models are expressed as power functions. Where spatial correlation is concerned, fractal systems have no characteristic scales; while where spatial entropy is concerned, fractal dimension just represents the characteristic value of entropy (Chen, 2017). This suggests that if entropy values depend on the scale of spatial measurement, we can convert the entropy values into fractal dimension values to avoid the scale dependence. Based on Renyi entropy, we can obtain multifractal parameter spectrums. This implies that we can utilize multifractal parameters to characterize the spatial heterogeneity of cities. This paper is devoted to researching the process of converting entropy measurement to fractal dimension for the scale-free spatial analysis of fractal urban phenomena. The rest parts are organized as follows. In Section 2, the relationships between entropy and fractal dimension are illustrated from the views of scale dependence and spatial heterogeneity. In Section [3](), an empirical analysis are made by means of the city of Beijing, the national capital of China, to verify the theoretical inferences. In [Section 4](), several related questions are discussed, and finally, the discussion are concluded by summarizing the main point of this work.

## 2. Theoretical models

### 2.1 Generalized entropy and fractal dimension

In a regular fractal, the complete parts which are similar to the whole are termed fractal units. In literature, fractal units are also called fractal copies ([Vicsek, 1989]()). A fractal system is a hierarchy of infinite levels with cascade structure. A fractal unit is a fractal subset or fractal subsystem at a given level. Fractal structure bear no characteristic scale, and cannot be described with the conventional measures such as length, area, and volume. In other words, the common measures of a fractal system depend heavily on the scales of measurement. The effective measurement of describing fractals is fractal dimension. To understand fractals, we must clarify the three properties of fractal systems: scaling law, fractal dimension, and entropy conservation.

**First, fractal systems follow the scaling law.** Scaling relation can be expressed as a functional equation as below ([Mandelbrot, 1982; Liu and Liu, 1993]()):

$$\mathrm{T}f(x) = f(\lambda x) = \lambda^b f(x), \tag{1}$$

where $f(x)$ represents a function of variable $x$, $\mathbf{T}$ denotes a dilation-contraction transform (scaling transform), $\lambda$ refers to scale factors, and $b$ to the scaling exponent. In mathematics, if a transform $\mathbf{T}$ is applied to a function $f(x)$, and the result is the function $f(x)$ multiplied by a constant $C$ (e.g., $C=\lambda^b$), then we will say that the function $f(x)$ is the eigenfunction under the transform $\mathbf{T}$, and the constant $C$ is the corresponding eigenvalue. This implies that a fractal model is just an eigenfunction of scaling transform, and the fractal dimension is associated with the eigenvalue $\lambda^b$. The solution to the [equation (1)]() is always a power function. Thus, a fractal is often formulated by a power law.

**Second, fractal systems bears fractal dimension.** A fractal dimension is usually a fractional dimension greater than its topological dimension. In Euclidean geometry, a point has 0 dimension, a line has 1 dimension, a plane has 2 dimensions, and a body has 3 dimensions. However, generally speaking, a fractal object cannot be characterized by the integer dimension. In many cases, the integer dimension is replaced by a fractional dimension which comes between 0 and 3. The fractal dimension of a geometric object is defined as a dimension that is strictly greater than the topological dimension of the object (Mandelbrot, 1982). Satisfying this condition, an integer dimension can also be treated as a fractal dimension. In fact, the dimensions of the space-filling curves such as Peano's curve and Hilbert's curve equal the Euclidean dimension of the corresponding embedding space, $d_E$. According to equation (1), fractal dimension can be defined by the scaling exponent $b$. Lets' see a simple fractal dimension

$$N(r) = N_1 r^{-D}, \tag{2}$$

in which $r$ is the scale of measurement, e.g., the linear size of boxes, $N(r)$ is the number of fractal copies based on the scale $r$, e.g., the number of nonempty boxes, $N_1$ refers to the proportionality coefficient, and $D$ to the fractal dimension. Apparently, equation (2) has invariance under the scaling transform, that is

$$N(\lambda r) = N_1(\lambda r)^{-D} = \lambda^{-D} N_1 r^{-D} = \lambda^{-D} N(r). \tag{3}$$

This indicates that the fractal model is the eigenfunction of the scaling transform, and the corresponding eigenvalue $C=\lambda^{-D}$ suggests the fractal dimension $D$, which is equivalent to the minus value of $b$ in equation (1). Based on box-counting method, the fractal parameter satisfy the following condition

$$d_T < D < d_E, \tag{4}$$

where $d_T$ refers to the topological dimension of a fractal object, and $d_E$ to the Euclidean dimension of the embedding space in which the fractal object exists.

**Third, fractal systems follow the law of entropy conservation.** Fractal systems can be described by a transcendental equation as follows

$$\sum_{i=1}^{N(r)} P_i(r)^q r_i^{(1-q)D_q} = 1, \tag{5}$$

where $P_i$ is the growth probability of the $i$th fractal unit, $r_i$ is the linear size of the $i$th fractal unit, $q$ denotes the order of moment, and the exponent $D_q$ represents the generalized correlation dimension. For a monofractal, i.e., a simple self-similar fractal, we have, $D_q \equiv D_0$; for a self-affine fractal, different directions have different fractal dimension values and for a given direction, we have $D_q \equiv D_0$. However, for a multifractal system, thing is complex. Different parts of a multifractal system have different characters, and can be described with different fractal dimension values. To simplify the process of spatial measurement, the varied linear scales $r_i$ can be substituted with a unified scale $r$. Thus, equation (5) can be re-written as

$$D_q = -\frac{M_q(r)}{\ln r} = \frac{1}{q-1}\frac{\ln \sum_{i=1}^{N(r)} P_i(r)^q}{\ln r}, \tag{6}$$

where $M_q$ denotes the generalized information entropy, namely, Renyi entropy. In fact, the generalized correlation dimension can also be termed generalized information dimension. The property of entropy conservation of a fractal system will be specially illustrated next. The Renyi entropy can be expressed as follows

$$M_q(r) = \frac{1}{1-q}\ln \sum_{i=1}^{N(r)} P_i(r)^q = -D_q \ln r, \tag{7}$$

which suggests that the generalized correlation dimension is just the characteristic value of Renyi entropy based on spatial scales (Chen, 2017). The generalized correlation dimension can be transformed into the mass exponent as below (Feder, 1988; Vicsek, 1989):

$$\tau_q = (q-1)D_q = \frac{(1-q)M_q(r)}{\ln r}, \tag{8}$$

where $\tau_q$ is termed the mass exponent of multifractal structure. Equation (8) shows the relationships between the generalized correlation dimension, the mass exponent and Renyi entropy. The generalized correlation dimension and the mass exponent compose the global parameter pair of multifractal analysis. The two parameters can be converted into a pair of local parameters of multifractals by Legendre's transform

$$\alpha(q) = \frac{d\tau(q)}{dq} = D_q + (q-1)\frac{dD_q}{dq}, \tag{9}$$

$$f(\alpha) = q\alpha(q) - \tau(q) = q\alpha(q) - (q-1)D_q. \tag{10}$$

where $f(\alpha)$ refers to the fractal dimension of the fractal units of certain sizes, and $\alpha(q)$ is the corresponding singularity exponent (Feder, 1988; Stanley and Meakin, 1988). If $D_q$ is termed global dimension of multifractal sets, then $f(\alpha)$ can be termed local dimension of the multifractals.

## 2.2 Scale dependence and entropy conservation of fractal urban systems

Global multifractal parameters are defined on the base of the scaling relation between Renyi entropy and the corresponding measurement scales. The parameter values of a multifractal systems such as cities based on a given approach (e.g., box-counting method) depend on the scope of study area (size, central location). That the size of study area influence fractal dimension estimation is a problem, but that the central location of study area influence fractal dimension measurement is an advantage rather than a problem. In fact, the similarity or even the commonality between entropy and fractal dimension lies in that both the entropy values and fractal dimension values depend on the method and study area. The advantage of entropy over fractal dimension is that entropy can be applied to measuring both Euclidean structure and fractal structure, while fractal dimension can only be applied to characterizing fractal structure (Chen and Feng, 2017). Compared with entropy, fractal dimension has two advantages. One is that fractal dimension values do not depend on the scale of measurement, the other is that fractal dimension values can reflect the local feature of random multifractals. The basic property of fractals (monofractal & multifractals) is that entropy conservation, that is, for a given level of a fractal hierarchy, different fractal units have the same entropy value. The entropy values of the fractal units at a given level in a fractal system depend on the growth probability distribution but are independent of spatial scales. This implies that entropy value cannot be used to describe the local features of different parts of a multifractal system of cities. In other words, entropy cannot reflect the spatial heterogeneity of a complex system. However, different fractal units have different fractal dimension values, which depend on both the growth probability distribution and spatial scales.

For random multifractals such as cities, which are in fact pre-fractals, we cannot identify entire fractal units, thus, both entropy and fractal dimension depend on the size and central location of study area. As we know, the entropy values of a system rely on two factors: one is the number of elements ($N$), and the other is the uniformity or homogeneity of the elements distribution. The size distribution of elements is reflected by probability structure, i.e., the difference of $P_i$ values. For a homogeneous system (say, a regular monofractal object), enlarge the size of study area, the entropy value will increase, but the location has no significant influence on the result; for a heterogeneous system (say, a random multifractal object), both the size and location of study area will impact on the entropy values: different area sizes indicate different element number ($N$), and different locations imply different probability distribution patterns of elements ($P_i$).

It is easy to demonstrate that the entropy values of a monofractal system depends on size of study area or scale of measurement. Let's see two simple examples, which are based on a regular fractal (Figure 1). The fractal was put forward by Jullien and Botet (1987) to reflect fractal growth and became well known due to the work of Vicsek (1989). So it was termed Vicsek's fractal, representing an embodiment of Stigler's law of eponymy (Stigler, 1980). This growing fractal was often employed to act as a simple fractal model of urban growth (Batty and Longley, 1994; Chen, 2012a; Frankhauser, 1998; Longley *et al*, 1991; White and Engelen, 1993). **(1) Entropy value depends on size of study area.** Please see the following regular growing fractal (Figure 1(a)). The first four steps represent a process of a growing prefractal. Different step reflects different size of study area. The first step is special and the results are outliers. You can see that the entropy values depend on the study area, but the fractal dimension value is certain. From the second step on, the entropy values and fractal dimensions are listed as below: **Step 1**: entropy $H=0$; fractal dimension $D=0$. For a point, the fractal dimension value can be obtained by L'Hospital's rule. **Step 2**: entropy $H = \ln(5) =1.6094$ nat; fractal dimension $D =- \ln(5)/\ln(1/3)=1.465$. **Step 3**: entropy $H = \ln(25) = 3.2189$ nat; fractal dimension $D = -\ln(25)/\ln(1/9)=1.465$. **Step 4**: entropy $H =\ln(125)=4.8283$ nat; fractal dimension $D =- \ln(125)/\ln(1/27) = 1.465$…. **(2) Entropy value also depends on scale of measurement.** Now, let's see the following regular growing fractal (Figure 1(b)). For this figure, different step reflects different linear scale of measurement. The first step is special and the results are outliers, too. The entropy values depend on the linear size, but the fractal dimension value is still certain. The entropy values and fractal dimensions are listed as below: **Step 1**: entropy $H=0$; fractal dimension $D=2$. For a surface, the fractal dimension can be obtained by L'Hospital's rule. **Step 2**: entropy $H = \ln(5) =1.6094$ nat; fractal dimension $D =- \ln(5)/\ln(1/3) =1.465$. **Step 3**: entropy $H =\ln(25) = 3.2189$ nat; fractal dimension $D = -\ln(25)/\ln(1/9) =1.465$. **Step 4**: entropy $H =\ln(125)=4.8283$ nat; fractal dimension $D =- \ln(125)/\ln(1/27) =1.465$…. For different fractal units in a given level (step), entropy value and fractal dimension value are both certain, that is, they are constant values (Table 1).

**Table 1 The values of entropy and fractal dimension of a regular growing monofractal system**

| Step for fractal generation | Figure 1(a): Variable size of study area and measurement scale | | Figure 1(b): Fixed size of study area and variable measurement scale | |
|---|---|---|---|---|
| | Entropy (nat) | Fractal dimension | Entropy (nat) | Fractal dimension |
| 1 (outlier) | 0 | 0 | 0 | 2 |
| 2 | 1.6094 | 1.4650 | 1.6094 | 1.4650 |
| 3 | 3.2189 | 1.4650 | 3.2189 | 1.4650 |
| 4 | 4.8283 | 1.4650 | 4.8283 | 1.4650 |
| … | … | … | … | … |
| $m$ | $\ln(5^{m-1})$ | $\ln(5^{m-1})/\ln(3^{m-1})$ | $\ln(5^{m-1})$ | $\ln(5^{m-1})/\ln(3^{m-1})$ |

**Note:** Different steps reflect different levels in a fractal hierarchy.

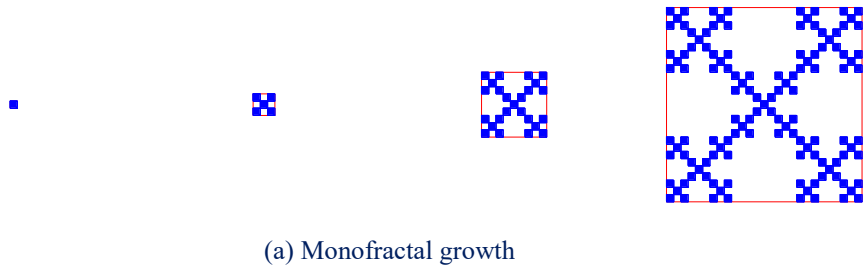

(a) Monofractal growth

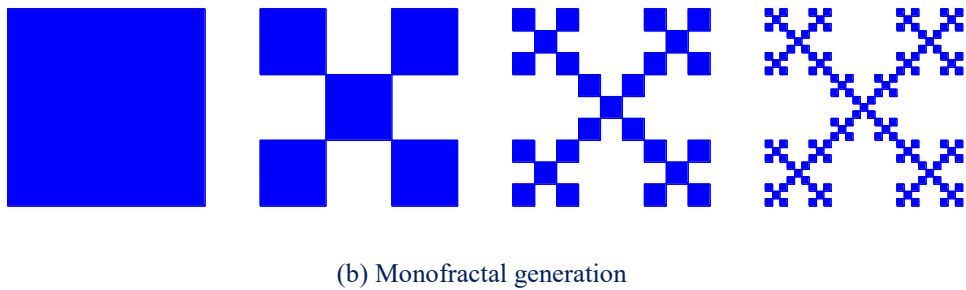

(b) Monofractal generation

**Figure 1 A regular growing monofractal which bears analogy with urban growth**

(**Note:** A monofractal possesses only one scaling process and is also termed "unifractal" in literature. Figure 1(a) represents the variable scale of measurement based on variable size of study area, and Figure 1(b) represents the variable scale of measurement based on fixed size of study area.)

The spatial structure of multifractal systems is different from that of simple fractal systems. For the multifractal systems, entropy values depend on size, location of study area as well as scale of measurement. Let's see an example of spatial heterogeneity and entropy conservation of multifractals. The following regular growing multifractals is well known for many fractal scientists and some urban geographers (Figure 2). The first step is special and the results are outliers, too. The entropy value depends on the linear size, but the box fractal dimension value is certain. From the second step on, the entropy values and fractal dimensions are listed as below: **Step 1**: entropy $H$=0; fractal dimension $D$ =0. **Step 2**: entropy $H$ = -ln(1/17)/17-4*4*ln(4/17)/17=1.5285 nat; box

dimension $D = -\ln(17)/\ln(1/5)=1.7604$. **Step 3**: entropy $H = -\ln(1/289)/289 - 8*4*\ln(4/289)/289 - 16*16*\ln(16/289)/289 = 3.0569$ nat; box dimension $D = -\ln(289)/\ln(1/25)=1.7604$. However, for different fractal units, entropy values are constant, but fractal dimension are different. In fact, for a multifractal object, different parts have different local fractal dimensions. The first three steps represent a multi-scaling prefractal. For example, for the second level of the third step, the five parts have two fractal dimension values. The central part, box dimension is $D=\ln(17/289)/\ln(2/25)=1.7604$; the other four parts, box dimension is $D=\ln(68/289)/\ln(10/25)=1.5791$. However, different parts have the same entropy values: entropy $H= -\ln(1/17)/17-4*4*\ln(4/17)/17=1.5285$ nat (Table 2).

Table 2 The values of entropy and fractal dimension of a regular growing multifractal systems

| Step for fractal generation | Global feature | | Local features | | | |
|---|---|---|---|---|---|---|
| | | | Central part | | Peripheral parts | |
| | Entropy (nat) | Fractal dimension | Entropy (nat) | Fractal dimension | Entropy (nat) | Fractal dimension |
| **1 (outlier)** | 0 | 0 | 0 | - | 0 | - |
| 2 | 1.5285 | 1.7604 | 0.1667 | 1.7604 | 1.3618 | 1.5791 |
| 3 | 3.0569 | 1.7604 | 1.5285 | 1.7604 | 1.5285 | 1.5791 |
| … | … | … | … | … | | |

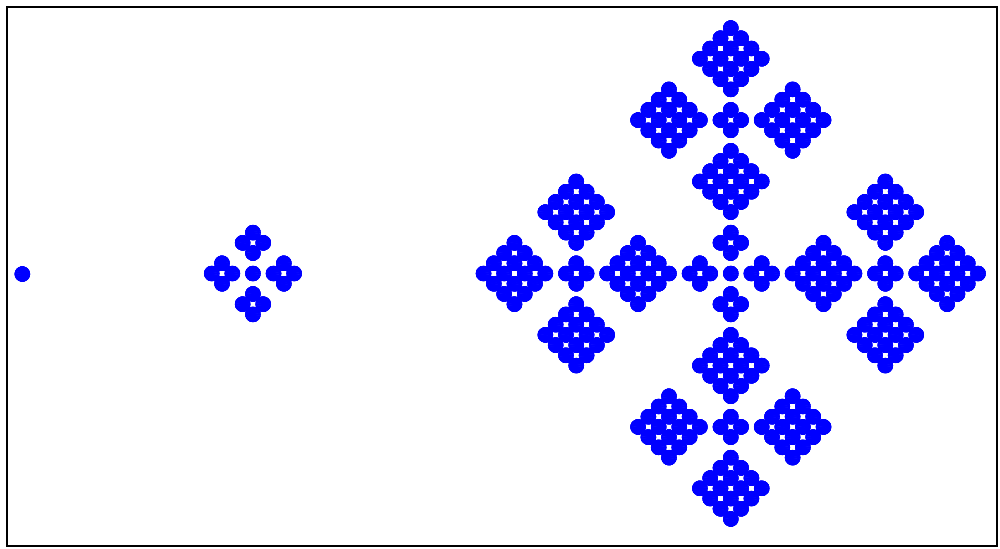

Figure 2 A regular growing multifractals which bears analogy with urban growth
(**Note:** To illustrate multifractal, Vicsek (1989) proposed this fractal with two different scales in the generator.)

## 2.3 Entropy-based fractal dimension analysis

According to the above analysis based on regular fractals, we can find two properties of fractal systems. **First, the entropy value of a fractal system depends on the scale of measurement, but the fractal dimension is independent of the scales**. For both simple fractals and multifractals, different steps represent different measurement scales. For monofractals, based on certain method, fractal dimension value is unique. However, for multifractals, different parts have different fractal dimension values. In contrast, for a given part of a multifractal system, the fractal dimension value does not depend on the measurement scales. **Second, different fractal units share the same entropy value**. The structure of a simple fractal is homogenous, and a fractal unit is the same as the other fractal unit. The entropy value of each fractal unit is the same. On the contrary, the structure of multifractals is heterogeneous, and a fractal unit may be different from another fractal unit. Despite the difference between fractal units, the entropy value of each fractal unit is still the same. However, different fractal units may have different fractal dimension values. This indicates that the fractal dimension of parts does not depend on measurement scales, but relies on local structure. Therefore, we can substitute fractal dimension for entropy to make spatial analysis of cities if one of the following two cases appears. One is that the measurement results depend on scales, and the other is that spatial heterogeneity must be taken into consideration.

In urban studies, it is convenient to transform spatial entropy into multifractal spectrums. The process is as follows: **(1) Transform Renyi entropy $M_q$ into global correlation dimension $D_q$ and mass exponent $\tau_q$**. It is easy to define global multifractal dimension based on Renyi entropy, which are applied to global spatial analyses. See equations (6) and (8). The global parameters comprise the generalized correlation dimension and mass exponent. **(2) Convert the global parameters into local multifractal parameters by Legendre transform**. See equations (9) and (10). The local parameters, including the local fractal dimension $f(\alpha)$ and the corresponding singularity exponent $\alpha(q)$, can be used to make partial spatial analysis. **(3) Substitute the spatial analysis by moment order analysis**. In practice, it is difficult to distinguish different spatial units of a random multifractal object from one another. A clever solution is to use moment analysis to replace local analysis. Map the parameter information of different spatial units into different orders of moment, $q$, thus we have multifractal parameter spectrums. A multifractal spectrum based on moment orders can be treated as the result of local scanning and sorting for a complex system (Chen, 2016; Huang and Chen, 2018).

## 3. Empirical analysis

### 3.1 Study area and methods

In this section, we will apply entropy measures and fractal dimension to urban form and growth. Urban form can be reflected and represented by urban population distributions, urban land use patterns, urban transport networks, and so on. The study area of this work is the urban agglomerations of Beijing city, the national capital of China, and the researched object is urban land use. The datasets came from the remote sensing images of four years, that is, 1984, 1994, 2006, and 2015 (Figure 3). A number of remote sensing images of Beijing from National Aeronautics and Space Administration (NASA) are available for spatial analysis. The ground resolution of these images is 30 meters (Chen and Wang, 2013). The functional box-counting method can be employed to measure the Renyi entropy and calculate multifractal parameters (Figure 4). This method was originally proposed by Lovejoy et al (1987) to estimate the fractal dimension of radar rain distribution. Later, Chen (1995) improved the method and used it to measure the fractal dimension of urban systems. The original functional box-counting method is based on the largest box with arbitrary area (Lovejoy et al, 1987), while the improved functional box-counting method is based on the largest box with a measure area of an urban envelope (Chen, 1995). This improved method is also termed Rectangle Space Subdivision (RSS) method (Chen and Wang, 2013; Feng and Chen, 2010). Where the studies on fractal cities are concerned, the improved functional box-counting method bears firm theoretical basis. On the one hand, its geometrical basis of RSS is the recursive subdivision of space and the cascade structure of hierarchies (Batty and Longley, 1994; Goodchild and Mark, 1987); on the other, its mathematical basis is the transformation relation between the power laws based on dilation symmetry and the exponential laws based on translational symmetry (Chen, 2012b).

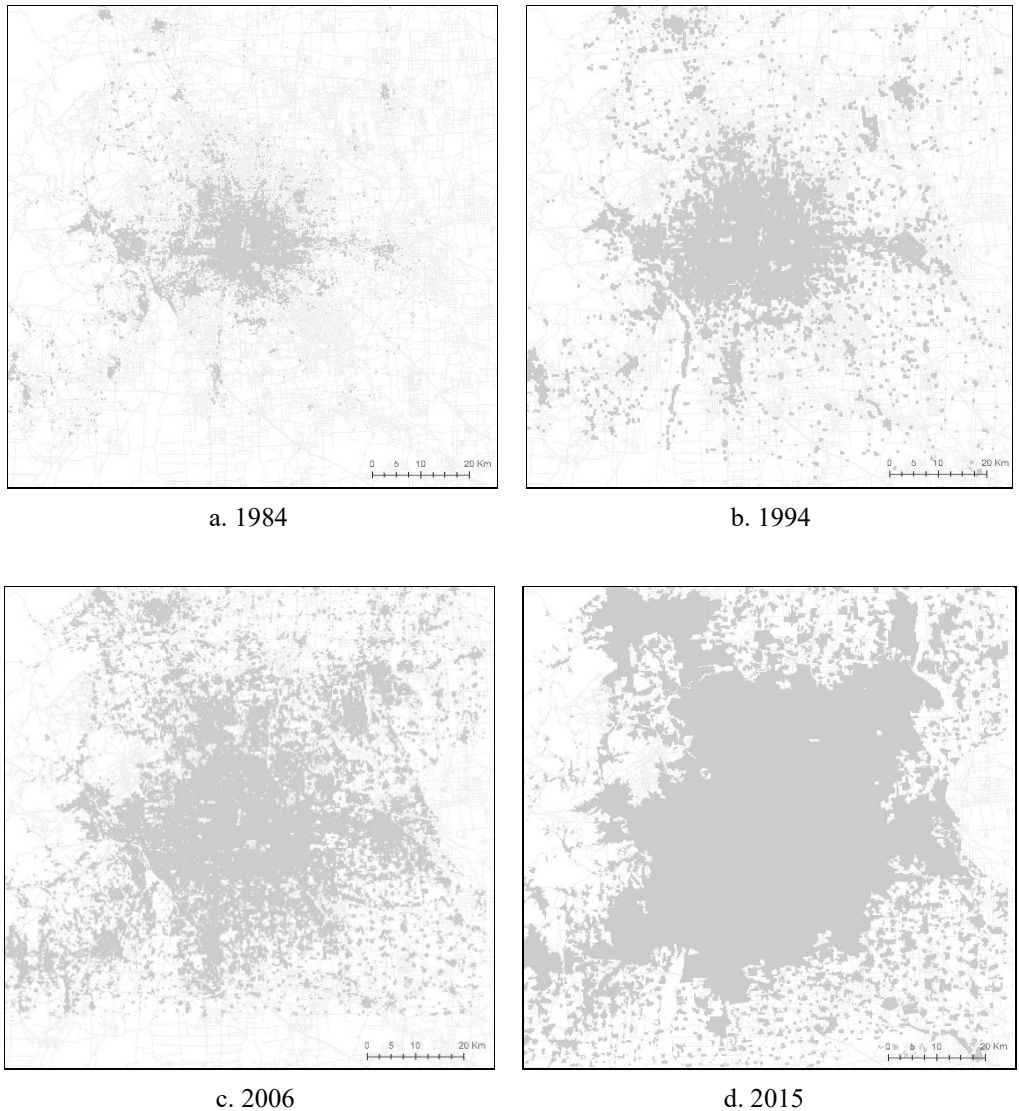

a. 1984  b. 1994

c. 2006  d. 2015

**Figure 3 Four typical images of Beijing's urban land use patterns**

The procedure of data extraction and parameter estimation comprises four steps. **Step 1: defining an urban boundary based on the recent image**. The most recent material we used was the remote sensing image of 2015. Based on this image, the boundary of Beijing city can be identified by using the "City Clustering Algorithm" (CCA) developed by Rozenfeld *et al* (2008, 2011). The urban boundary can be termed as urban envelope (Batty and Longley, 1994; Longley *et al*, 1991). Then, a measure area can be determined in terms of the urban envelope (Chen *et al*, 2017). **Step 2: extracting the spatial dataset using the function box-counting method**. First of all, we can extract the dataset from the image of the recent year (2015). A set of boxes is actually a grid of rectangular squares, each of which has an area of urban land use. The area may be represented by the pixel number. Therefore, in the dataset, each number represents a value of land use area of the urban pattern falling into a box (square). Changing the linear size of the boxes, we will have different dataset. The box system forms a hierarchy of grids, which yield a hierarchy of spatial datasets. Applying the system of boxes to the images in different years, we have different datasets for calculating spatial entropy and fractal dimension. **Step 3: calculating the spatial Renyi entropy**. Using equation (7), we can compute the Renyi entropy of urban land use based on given linear size of functional boxes. For each linear size of boxes, we can obtain an entropy value for Beijing's urban form. For each year, we have a number of sets of entropy values based on different linear sizes of boxes. If the entropy values based on different box sizes have no significant differences, we can utilize the means of Renyi entropy values to make spatial analysis of urban form and growth. **Step 4: computing the multifractal parameter spectrums**. If the entropy values depend heavily on the linear sizes of boxes, we should transform the Renyi entropy into the generalized correlation dimension using equations (6) and (7). For different linear sizes of boxes $r$, we have different Renyi entropy values $M_q(r)$. As shown by equation (7), there is a linear relation between $\ln(r)$ and $M_q(r)$. The analytical process can be illustrated as follows (Figure 5).

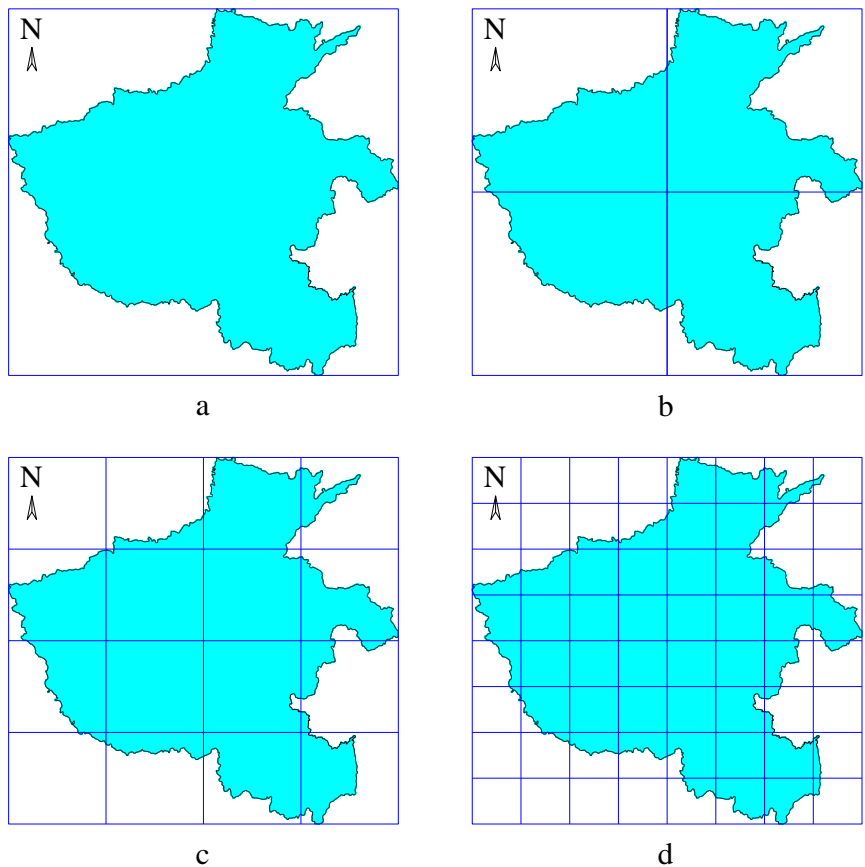

**Figure 4 A sketch map of the functional box-counting method for spatial entropy and fractal**

dimension measurement (the first four steps)

(**Note:** The closed urban boundary curve is termed *urban envelope*, and the rectangle hugging closely the urban envelope gives a *measure area* of the urban envelope. The grid represents the functional boxes.)

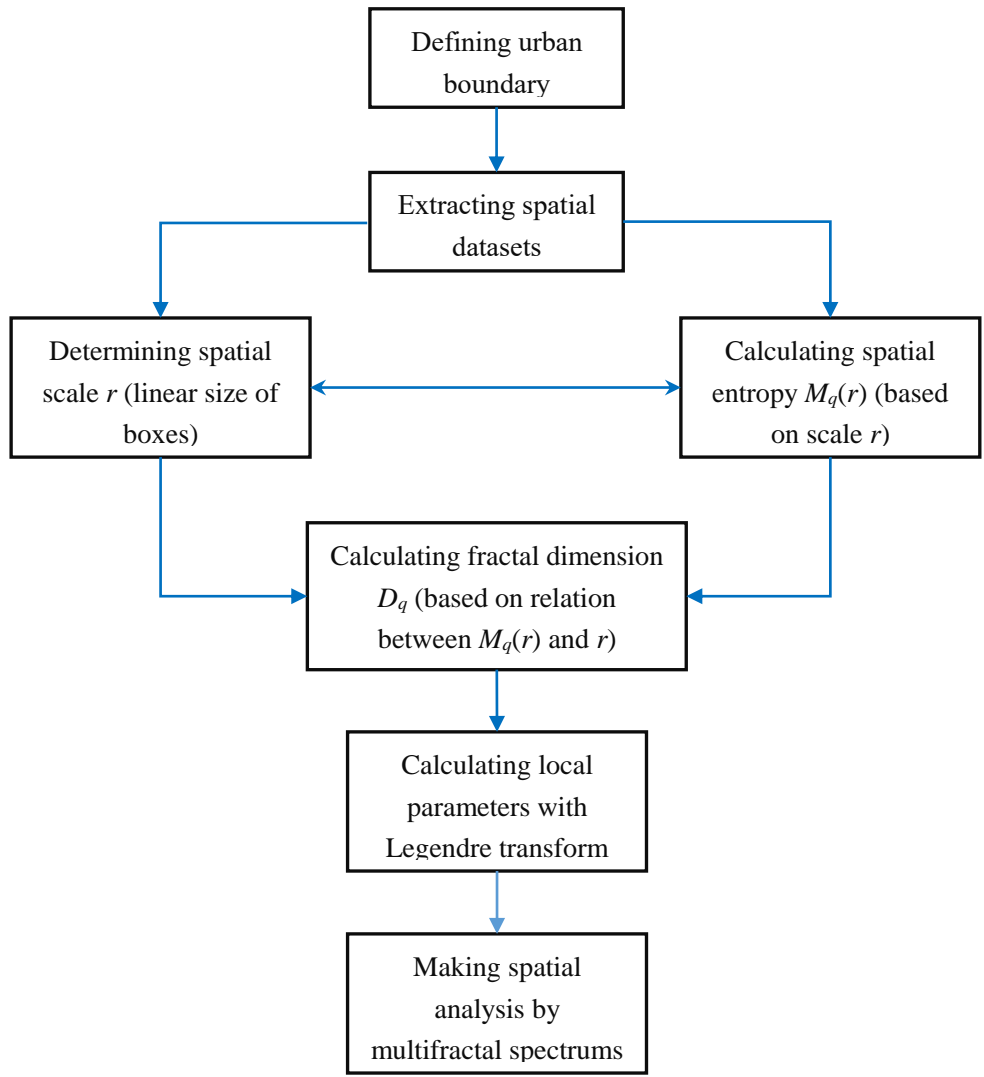

**Figure 5 A flow chart of spatial analysis for cities from spatial entropy to multifractal spectrums**
(**Note:** Spatial entropy can be used to make spatial analysis of cities based on characteristic scales, while multifractal spectrums can be employed to make spatial analysis based on scaling in cities.)

   The process of parameter estimation is simple by means of the least square calculations. Using linear regression technique, we can estimate the generalized correlation dimension $D_q$, which is just the slope of the semi-logarithmic equation. It should be noted that the regression equation has no intercept (Huang and Chen, 2018). If $q=1$, equation (7) will be invalid. In this case, according to the well-known L'Hospitale rule, the Renyi entropy will be replaced by the Shannon entropy, which

can be expressed as

$$H(r) = M_1(r) = -\sum_{i=1}^{N(r)} P_i(r) \ln P_i(r), \quad (11)$$

where $H(r)$ denotes Shannon's information entropy based on the linear size of boxes $r$. This implies that the Shannon entropy is the special case of the Renyi entropy. Apply Shannon entropy to geographical analysis yield the important concept of spatial entropy (Batty, 2010). In fact, Renyi's entropy can be regarded as the generalization of Shannon's entropy. In short, for $q=1$, equation (7) will be substituted by the following relation

$$H(r) = -\sum_{i=1}^{N(r)} P_i(r) \ln P_i(r) = -D_1 \ln r, \quad (12)$$

which will give the information dimension of the multifractal dimension spectrums.

### 3.2 Results and findings

The above process of data extraction and parameter estimation is convenient by means of ArcGIS technique and mathematical computation software such as Matlab. Partial spatial Renyi entropy for Beijing are shown in Table 3, and the corresponding multifractal parameters are displayed in Table 4. More results can be found in the attached files of Excel data. If the moment order $q=0$, we have Boltzmann macro state entropy; If $q=1$, we have Shannon information entropy; If $q=2$, we have Renyi correlation entropy. For arbitrary order of moment $q$, we have Renyi's generalized entropy. Obviously, for a given order of moment, say, $q=0$, the entropy $M_0(r)$ value depends significantly on the linear sizes of boxes $r$ (Figure 6, Table 3). In other words, the spatial Renyi entropy values of Beijing urban land use rely on the scales of measurement. Based on different linear sizes of boxes, the entropy values are different. In particular, the average value of the spatial entropy are invalid because the mean depends on the size of datasets. That is to say, changing the range of the linear sizes of boxes yields different average values of Renyi entropy.

Table 3 Partial generalized entropy values of Beijing's urban land use pattern in 2015

| Moment order $q$ | Generalized entropy based on different scales $r$ | | | | | | | |
|---|---|---|---|---|---|---|---|---|
| | $r=1/2$ | $r=1/4$ | $r=1/8$ | $r=1/16$ | $r=1/32$ | $r=1/64$ | $r=1/128$ | $r=1/256$ |
| -20 | 1.8225 | 3.6449 | 5.4674 | 7.2898 | 9.1123 | 10.9347 | 12.7572 | 14.5797 |
| -15 | 1.8045 | 3.6089 | 5.4134 | 7.2179 | 9.0223 | 10.8268 | 12.6313 | 14.4357 |
| -10 | 1.7702 | 3.5405 | 5.3107 | 7.0809 | 8.8511 | 10.6214 | 12.3916 | 14.1618 |
| -5 | 1.6806 | 3.3612 | 5.0417 | 6.7223 | 8.4029 | 10.0835 | 11.7640 | 13.4446 |
| -4 | 1.6430 | 3.2859 | 4.9289 | 6.5718 | 8.2148 | 9.8578 | 11.5007 | 13.1437 |
| -3 | 1.5901 | 3.1801 | 4.7702 | 6.3602 | 7.9503 | 9.5403 | 11.1304 | 12.7204 |
| -2 | 1.5123 | 3.0246 | 4.5369 | 6.0493 | 7.5616 | 9.0739 | 10.5862 | 12.0985 |
| -1 | 1.4058 | 2.8116 | 4.2174 | 5.6232 | 7.0290 | 8.4348 | 9.8406 | 11.2464 |
| 0 | 1.3410 | 2.6820 | 4.0230 | 5.3640 | 6.7050 | 8.0460 | 9.3870 | 10.7280 |
| 1 | 1.3226 | 2.6452 | 3.9679 | 5.2905 | 6.6131 | 7.9357 | 9.2583 | 10.5809 |
| 2 | 1.3148 | 2.6296 | 3.9443 | 5.2591 | 6.5739 | 7.8887 | 9.2035 | 10.5183 |
| 3 | 1.3105 | 2.6209 | 3.9314 | 5.2419 | 6.5523 | 7.8628 | 9.1732 | 10.4837 |
| 4 | 1.3078 | 2.6155 | 3.9233 | 5.2311 | 6.5388 | 7.8466 | 9.1544 | 10.4621 |
| 5 | 1.3059 | 2.6119 | 3.9178 | 5.2237 | 6.5297 | 7.8356 | 9.1416 | 10.4475 |
| 10 | 1.3017 | 2.6035 | 3.9052 | 5.2069 | 6.5087 | 7.8104 | 9.1122 | 10.4139 |
| 15 | 1.3001 | 2.6003 | 3.9004 | 5.2006 | 6.5007 | 7.8008 | 9.1010 | 10.4011 |
| 20 | 1.2993 | 2.5986 | 3.8979 | 5.1972 | 6.4965 | 7.7957 | 9.0950 | 10.3943 |

**Note**: For $q=1$, the numbers represent Shannon's information entropy values.

If we convert the Renyi's entropy values into multifractal parameters, the value of a parameter is unique. For the moment order $q=0$, we can transform a series of Boltzmann macro state entropy $M_0(r)$ values into a capacity dimension $D_0$ value; For $q=1$, we can transform a series of Shannon information entropy $M_1(r)$ values into an information dimension $D_1$ value; For $q=2$, we can transform a series of Renyi correlation entropy $M_2(r)$ values into a correlation dimension $D_2$ value. For arbitrary order of moment $q$, we can transform Renyi's generalized entropy $M_q(r)$ values into a set of generalized correlation dimension $D_q$ values. Apparently, for given order of moment, say, $q=1$, the fractal dimension $D_1$ value is independent of the linear sizes of boxes $r$ (Figure 7). Using equation (8), we can convert the generalized correlation dimension $D_q$ values into the mass exponent $\tau_q$ values. The generalized correlation dimension $D_q$ and mass exponent $\tau_q$ belong to the global parameters of multifractal models. By means of Legendre transform, equations (9) and (10), we can transform the global parameters into local parameters, including the singularity exponent $\alpha(q)$ and the corresponding fractal dimension $f(\alpha(q))$ (Table 4). Based on the global parameters, we have the global multifractal spectrum, i.e., $D_q$-$q$ spectrums (Figure 7); based on the local parameters, we have the local multifractal spectrum, i.e., $f(\alpha)$-$\alpha$ spectrums (Figure 8). The latter is often termed $f(\alpha)$ curve in literature (Feder, 1988). In practice, we can compute the local parameter values by using the $\mu$-weight method first (Chhabra and Jensen, 1989; Chhabra et al, 1989). Then, using Legendre transform, we can converted the local parameter values into the global parameter values (Chen, 2014; Chen and Wang, 2013; Huang and Chen, 2018).

Table 4 Partial multifractal parameter values of Beijing's urban land use pattern in 2015

| Moment order $q$ | Fractal parameter and goodness of fit | | | | | | |
|---|---|---|---|---|---|---|---|
| | $D_q$ | $R^2$ | $\tau_q$ | $\alpha(q)$ | $R^2$ | $f(\alpha)$ | $R^2$ |
| -20 | 2.6293 | 0.8308 | -55.2143 | 2.7124 | 0.8113 | 0.9664 | 0.7092 |
| -15 | 2.6033 | 0.8369 | -41.6527 | 2.7122 | 0.8115 | 0.9698 | 0.7112 |
| -10 | 2.5539 | 0.8484 | -28.0929 | 2.7116 | 0.8120 | 0.9771 | 0.7159 |
| -5 | 2.4246 | 0.8790 | -14.5473 | 2.7015 | 0.8116 | 1.0397 | 0.7361 |
| -4 | 2.3703 | 0.8927 | -11.8515 | 2.6884 | 0.8082 | 1.0979 | 0.7423 |
| -3 | 2.2940 | 0.9137 | -9.1759 | 2.6592 | 0.8012 | 1.1983 | 0.7535 |
| -2 | 2.1818 | 0.9470 | -6.5454 | 2.5898 | 0.7984 | 1.3658 | 0.8074 |
| -1 | 2.0281 | 0.9874 | -4.0563 | 2.3288 | 0.8782 | 1.7275 | 0.9683 |
| 0 | 1.9346 | 0.9992 | -1.9346 | 1.9791 | 0.9951 | 1.9346 | 0.9992 |
| 1 | 1.9081 | 1.0000 | 0.0000 | 1.9081 | 1.0000 | 1.9081 | 1.0000 |
| 2 | 1.8968 | 0.9998 | 1.8968 | 1.8888 | 0.9994 | 1.8807 | 0.9987 |
| 3 | 1.8906 | 0.9995 | 3.7812 | 1.8810 | 0.9986 | 1.8618 | 0.9959 |
| 4 | 1.8867 | 0.9992 | 5.6601 | 1.8773 | 0.9982 | 1.8489 | 0.9927 |
| 5 | 1.8841 | 0.9989 | 7.5363 | 1.8752 | 0.9979 | 1.8399 | 0.9900 |
| 10 | 1.8780 | 0.9982 | 16.9021 | 1.8720 | 0.9974 | 1.8181 | 0.9824 |
| 15 | 1.8757 | 0.9979 | 26.2599 | 1.8712 | 0.9973 | 1.8086 | 0.9784 |
| 20 | 1.8745 | 0.9978 | 35.6151 | 1.8709 | 0.9972 | 1.8031 | 0.9757 |

**Note**: The global parameter values are estimated using equations (6), (7), and (8), while the local parameters are estimated by means of the $\mu$-weight method.

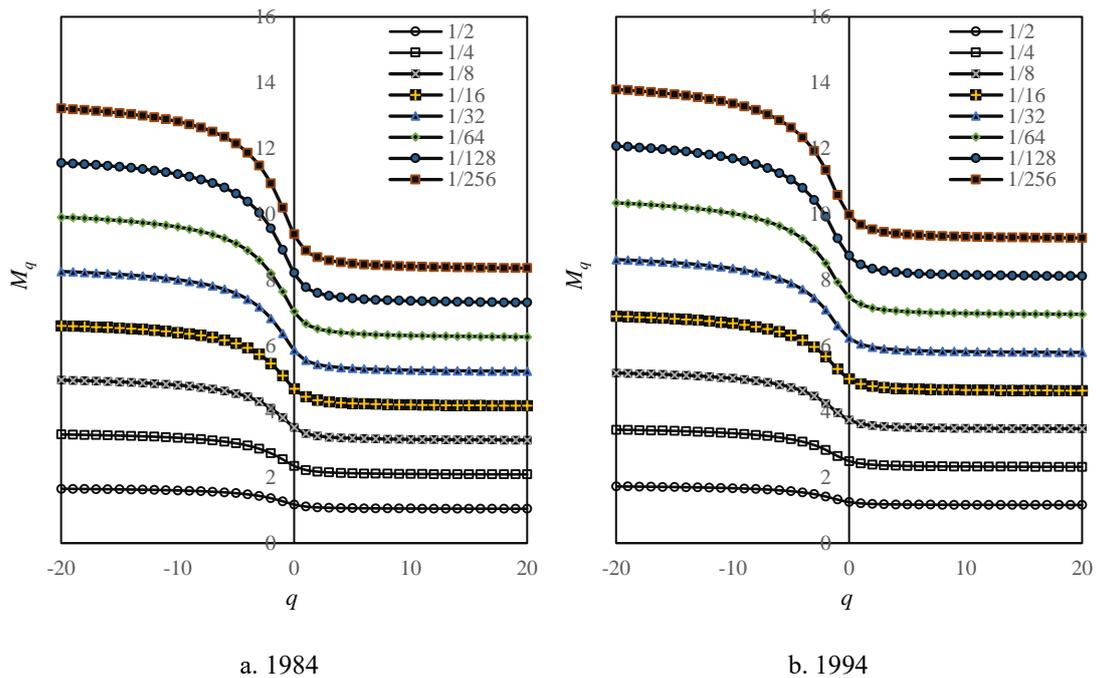

a. 1984　　　　　　　　　　　　　　b. 1994

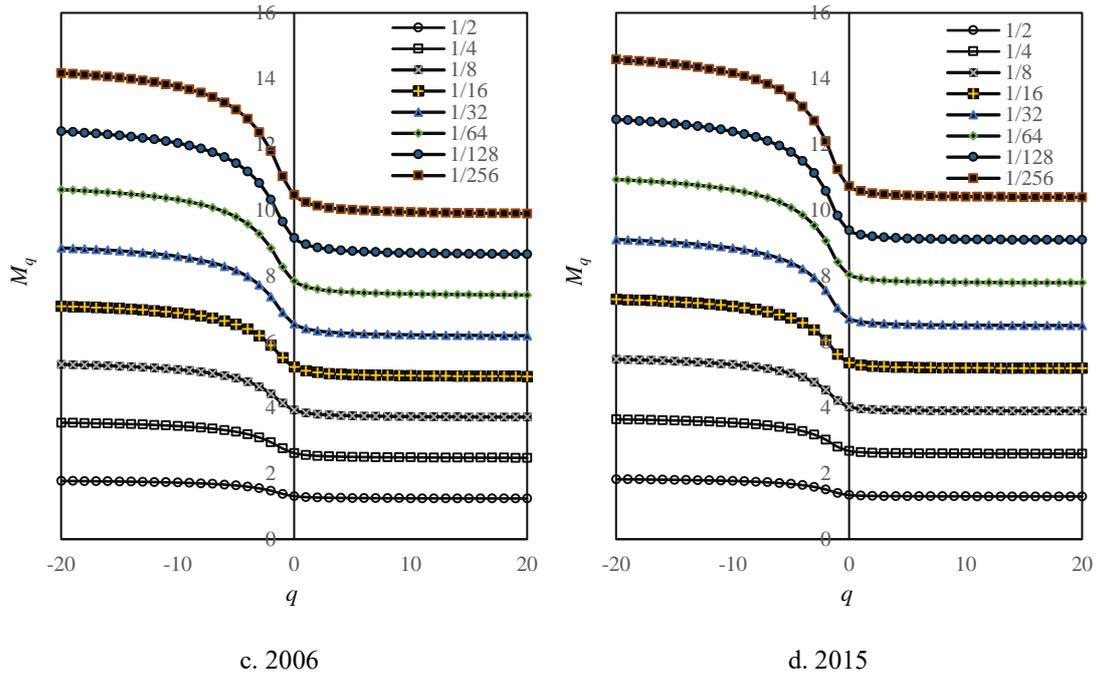

c. 2006                                    d. 2015

**Figure 6 The Renyi entropy spectrums based on moment order parameter and different spatial scales of measurement**

**(Note:** From the bottom to the top, the linear size of functional boxes are $r$=1/2, 1/4, 1/8, 1/16, 1/32, 1/64, 1/128, and 1/256, respectively. Different linear sizes of the boxes represents different spatial scales of measurements, and different Renyi entropy spectral lines based on different box sizes reflect the scale-dependence of spatial entropy.)

The main task of this article is not to explore the land use patterns of Beijing city. Instead, this paper is devoted to solving the problem of scale dependence of spatial entropy using fractal dimension. Nevertheless, we still discuss the growth characteristics of Beijing by means of complexity measures. It is difficult to make spatial analysis of urban form of Beijing using spatial Renyi entropy. Due to scale dependence of spatial measurements, the spectral curves of Renyi entropy are dazzling (Figure 6). In contrast, it is easy to make a spatial analysis using multifractal spectrums because there is only one spectral line for a given fractal parameter in a given year. The global multifractal parameters can be used to analyze the spatial correlation of urban evolution (Figure 6), while the local parameters can be employed to analyze the spatial heterogeneity of urban structure (Figure 7). Fractal dimension can be utilized to measure the space filling extent, spatial uniformity, and spatial complexity. According to the multifractal spectrums, the chief characteristics of Beijing's urban form and growth are as follows. **First, Beijing space filling speed was too fast, and space filing extent was too high**. From 1984 to 1994 to 2006 to 2015, the capacity dimension $D_0$ values increased from 1.6932 to 1.8011 to 1.8877 to 1.9346. By means of the formula $v=D_0/2$, we can calculate the space filling rate of urban form, *v*, and the results are 0.8466, 0.9005, 0.9439, and 0.9673. In recent years, the space filling rate is close to the upper limit 1. **Second, spatial heterogeneity became weaker and weaker**. From 1984 to 1994 to 2006 to 2015, the information dimension $D_1$ values went up from 1.6048 to 1.7467 to 1.8468 to 1.9081. Using the formula $u=1-D_1/2$, we can calculate the spatial redundancy rate of urban form, *u*, and the results are 0.1976, 0.1267, 0.0766, and 0.0459 (Table A). The spatial redundancy rate is in fact an index of spatial heterogeneity. Reduction of redundancy indicates weakening of spatial heterogeneity. **Third, the growth of Beijing city is of outward expansion**. The closer to the center area, the faster the space filling speed will be. In terms of local fractal spectrums, city development can be classified into two types: one is central aggregation, and the other is peripheral expansion (Chen, 2014). Beijing's city development belong to the former type (Figure 8). However, the space filling speed in the central area is obviously faster than that in the edge area (Figure 7). **Fourth, there was excessive correlation in urban fringe**. Generally speaking, the generalized correlation dimension value come between 0 and 2. However, when the order of moment *q* approaches to negative infinity, the $D_q$ values exceeded 2 and became bigger and bigger (Figure 7). This suggests that there are too many messy patches of land use to fill the urban fringe. **Fifth, the quality of spatial structure declined**. A local multifractal spectrum is supposed to be a smooth single-peak curve. In 1984, the local fractal dimension spectral lines is regular. However, from 1995 to 2015, the *f(α)* curves deviated more and more from the normative spectral line (Figure 8).

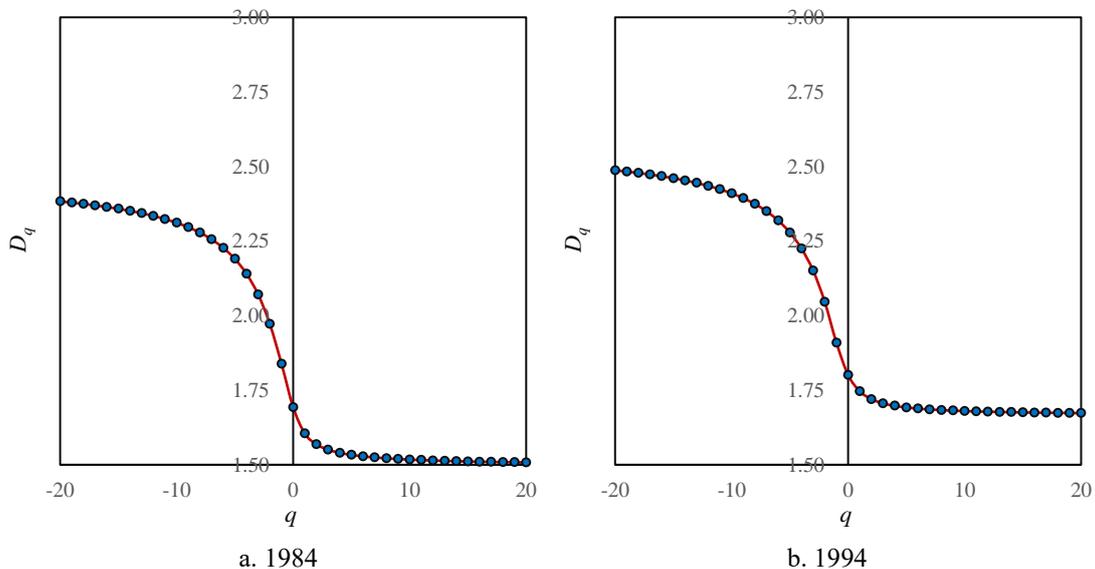

a. 1984　　　　　　　　　　　　　　b. 1994

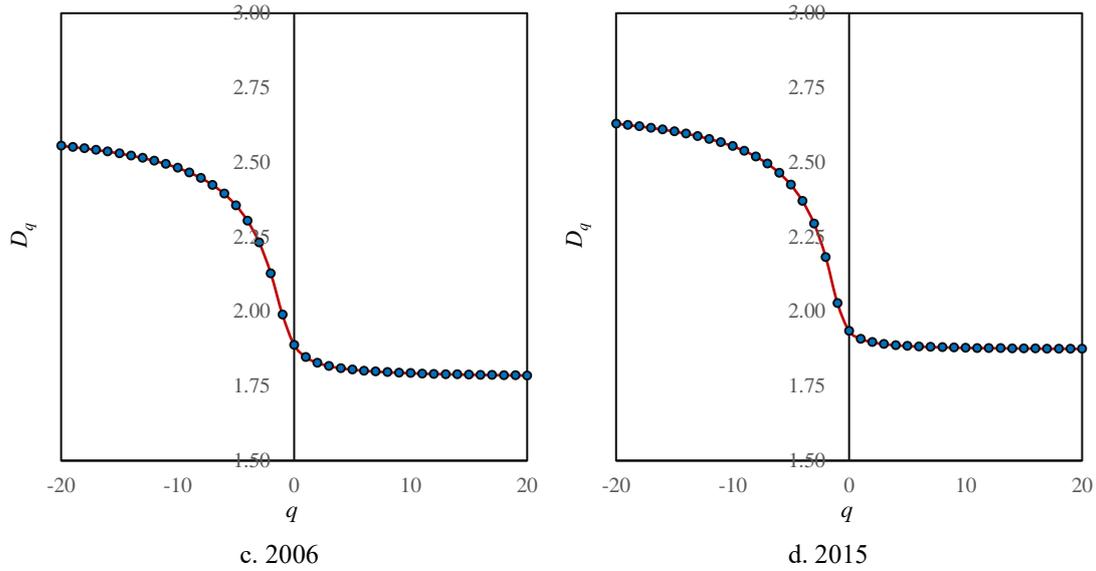

c. 2006　　　　　　　　　　　　　d. 2015

**Figure 7 The global multifractal spectrums based on moment order parameter**

(**Note:** Based on different linear sizes of functional boxes, i.e., $r$=1/2, 1/4, 1/8, 1/16, 1/32, 1/64, 1/128, and 1/256, and the corresponding Renyi entropy $M_q(r)$ values, the generalized correlation dimension $D_q$ can be evaluated. Multifractal parameter values depend on moment order $q$, but are independent of spatial scale $r$.)

## 4. Discussion

Entropy and fractal dimension are two important measures of spatial complexity in geographical world. Substituting spatial Renyi entropy by multifractal parameters, we can solve two problems for urban studies. One is the scale dependence of entropy measurement, and the other is the description of spatial heterogeneity of urban morphology. In particular, if we convert spatial entropy into fractal dimension, a number of entropy values based on different scales can be represented by one fractal dimension which is dependent of scales. Thus, many numbers are condensed into one number, so that the description and analytical process will become simpler (Table 5). These properties have been illustrated by the above case study of Beijing city. In fact, the fractal models can associate spatial correlation functions with entropy functions (Chen, 2014). Therefore, based on fractal dimension, the concept of scale dependence is replaced by the notion of spatial dependence. Spatial dependence (spatial correlation) and spatial heterogeneity (spatial difference) represent two essential aspects of geographical systems (Anselin, 1996; Goodchild, 2004). For a simple system, the spatial entropy has a determinate value. However, for a complex system such as cities, the values of spatial entropy depend on the scales of measurement, and thus we cannot find a certain entropy value for urban form and urban systems. It is advisable to transform spatial entropy into fractal parameters. On the other hand, multifractal scaling provides a quantitative characterization of heterogeneous phenomena (Stanley and Meakin, 1988). If we want to explore spatial heterogeneity deeply in a complex spatial system such as cities, the limitation of entropy will also appear. Due to entropy conservation, different parts of a fractal urban system bears the same entropy value. So, we cannot bring to light the local features by spatial entropy. In this case, we can use multifractal parameters to characterize the spatial heterogeneity of urban form and urban systems (Figure 9).

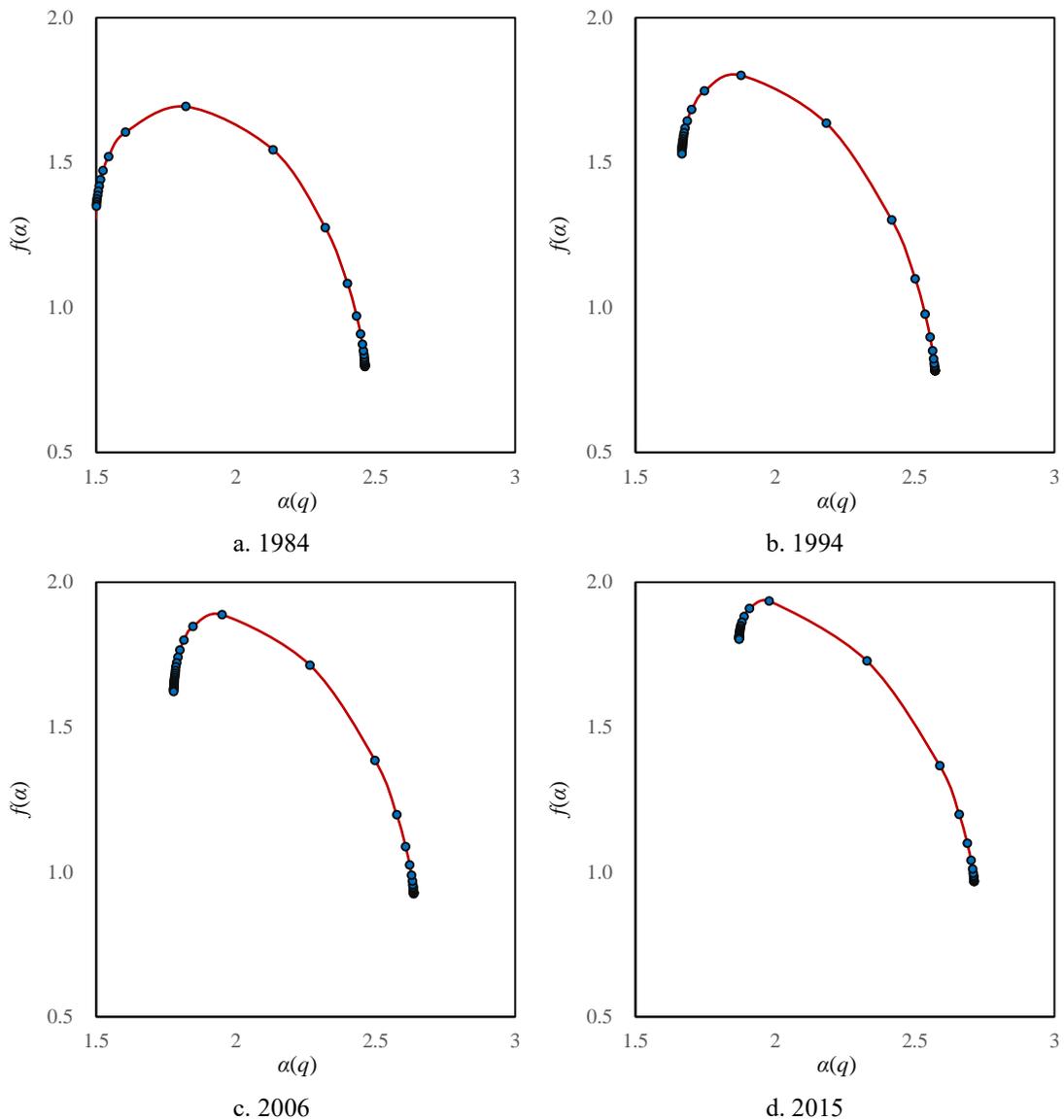

**Figure 8 The local multifractal spectrums based on singularity exponent, i.e., the $f(\alpha)$ curves**
(**Note:** A local multifractal spectrum is a unimodal curve, which can be used to reflect the aggregation or diffusion of a city's evolution)

Spatial entropy measure has a natural connection with fractal dimension of urban systems. In literature, both entropy and fractal dimension have been employed to characterize urban patterns and evolution process (e.g., Encarnação *et al*, 2013; Fan *et al*, 2017; Padmanaban *et al*, 2017; Terzi and Kaya, 2011). However, the scale dependence of spatial entropy and its relationship with fractal dimension are rarely reported. The scale dependence of spatial entropy measurement is associated with the scale-free property of urban systems. Fractal dimension can be used to act as the characteristic parameter of urban description. This problem has been preliminarily researched in previous works (Chen *et al*, 2017; Chen and Feng, 2017). In one companion paper, using box-counting method, we reveal that spatial entropy values depend on the scales of measurement and the normalized entropy values are empirically equal to the normalized fractal dimension values (Chen *et al*, 2017). This suggests that two approaches can be utilized to solve the problem of scale dependence of spatial entropy. One is to use fractal dimension to replace spatial entropy, and the other is to normalize spatial entropy. Three typical fractal dimensions in global multifractal dimension, i.e., capacity dimension, information dimension, and correlation dimension are discussed in this research, but the results have not been generalized to multifractal parameter spectrums. In another companion paper, based on area-radius scaling, the normalized Renyi entropy is generalized to multifractal spectrums (Chen and Feng, 2017). Two sets of multifractal indicators are proposed to describe urban growth and form. The mathematical modeling based on characteristic scales and the spatial analysis based on scaling are integrated into a logic framework. Compared with the previous studies, this work bears third new points. **First, the scale dependence of spatial Renyi entropy is illustrated by box-counting method**. Changing the linear sizes of boxes yields different entropy spectral curves. It is complicated to make spatial analysis of cities using these spectral curves. **Second, the solution to the scale dependence problem of spatial entropy is clarified**. Transforming the Renyi entropy into multifractal dimension, the different entropy values based on different measurement scales will be replaced by a fractal dimension value, which is actually a characteristic value of spatial entropy and independent of scales of measurement. **Third, similarities and differences between spatial entropy and fractal dimension spectrums are discussed.** Spatial entropy is simple and easy to understand, but it cannot be used to describe the spatial heterogeneity of city systems. In contrast, using multifractal parameter spectrums, we can characterize the spatial heterogeneity of urban form and urban systems. The main shortcomings of this work rest with three aspects. First, the empirical analysis is chiefly based on box-counting method. The other method such as Sandbox method, growing cluster method, and so on, are not taken into account for the time being. All these method can be applied to the studies on fractal cities. Second, the uncertainty of fractal dimension is not discussed. The fractal dimension values of urban form and urban systems depend on the size and central location of a study area.

**Table 5 A comparison of merits and demerits between spatial entropy and fractal dimension for spatial analysis of cities**

| Item | | Spatial entropy | Fractal dimension |
|---|---|---|---|
| **Similarity** | | Can be measured by box-counting method | Can be measured by box-counting method |
| **Difference** | Advantages | (1) Simplicity for computation and understanding; (2) Suitable for both the systems with characteristic scales and the scale-free systems | (1) Does not depend on scale of measurement; (2) Simplicity for analytical process; (3) Associate spatial dependence with spatial heterogeneity |
| | Disadvantages | (1) Scale dependence of measurement; (2) Complexity for analytical process; (3) Cannot describe spatial heterogeneity | (1) Complexity for computation and understanding; (2) Not suitable for the systems with characteristic scales |

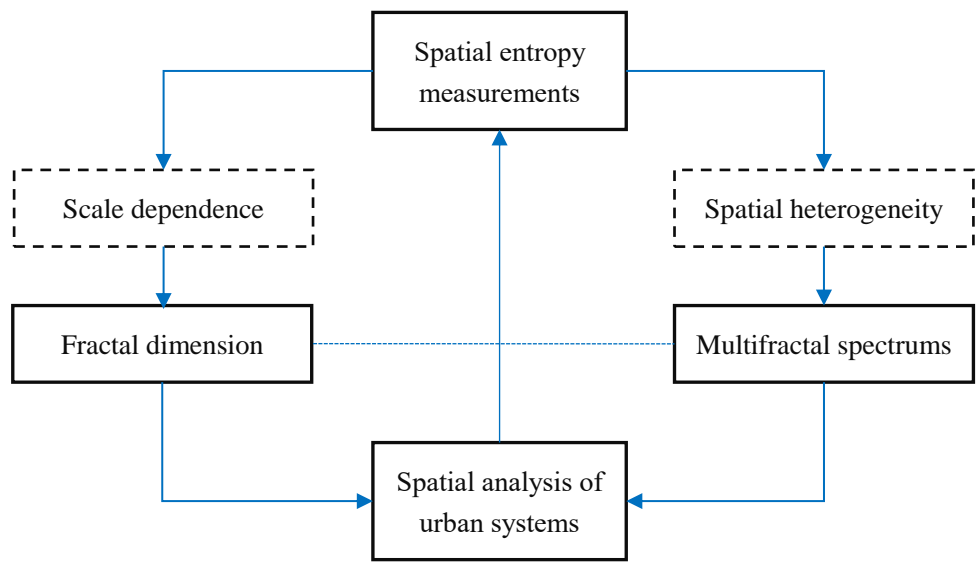

**Figure 9 Two cases of spatial entropy analyses transformed into fractal dimension analyses**
(**Note:** The problem of scale dependence of entropy measurement can be solved by transforming entropy into fractal dimension by means of scaling relation, and the property of spatial heterogeneity can be characterized by multifractal parameter spectrums.)

# 5. Conclusions

Fractal dimension is defined on the base of entropy function, and this suggests that the spatial entropy can be associated with fractal dimension of cities. Based on the theoretical exploration and empirical analysis, the main conclusions of this paper can be reached as follows. **First, fractal dimension can be used to solve the problem of scale dependence of spatial entropy of cities.** For the simple spatial systems, we can obtain determinate entropy values. However, for the complex spatial systems such as cities and systems of cities, we cannot gain certain entropy values. Both Shannon's information entropy and Renyi entropy spectrum depend on the scale of measurement. The uncertainty of entropy values give rise to trouble for spatial modeling and analysis of cities. One of effective method of solving the problem is to substitute the spatial entropy with fractal dimension. Fractal dimension values do not depend on the scales of measurement. We can use the capacity dimension to replace the macro state entropy, use the information dimension to replace Shannon's entropy, and use the generalized correlation dimension spectrum to replace Renyi's entropy spectrum. **Second, multifractal scaling can be employed to describe the spatial heterogeneity of cities.** The scale dependence indicates fractals and scaling. Simple fractal systems have homogeneous structure and different parts have the same entropy and fractal dimension. However, complex spatial systems such as cities and systems of cities have heterogeneous structure, and different parts have different local fractal dimension values, but have the same entropy value. This suggests that Renyi entropy values cannot reflect the spatial differences of complex spatial systems such as cities. In contrast, multifractal dimension spectrums can be used to reveal the spatial heterogeneity of complex systems, including urban form and urban systems. Among various multifractal parameter, the spatial redundancy rate based on information dimension can be used as a concise index of spatial heterogeneity of cities.

# Acknowledgements

This research was sponsored by the National Natural Science Foundation of China (Grant No.

41671167). The support is gratefully acknowledged.

# Appendix

**Table A The main fractal dimension values and the related spatial measurements of urban growth**

| Year | Capacity dimension $D_0$ | Information dimension $D_1$ | Correlation dimension $D_2$ | Spatial filling rate $v=D_0/D_{max}$ | Spatial redundancy rate $u=1-D_1/D_{max}$ |
|---|---|---|---|---|---|
| **1984** | 1.6932 | 1.6048 | 1.5682 | 0.8466 | 0.1976 |
| **1994** | 1.8011 | 1.7467 | 1.7199 | 0.9005 | 0.1267 |
| **2006** | 1.8877 | 1.8468 | 1.8277 | 0.9439 | 0.0766 |
| **2015** | 1.9346 | 1.9081 | 1.8968 | 0.9673 | 0.0459 |